\documentclass[]{pasj01}

\begin{document} 
\Received{}
\Accepted{}

\title{The Herschel-PACS North Ecliptic Pole Survey}

\author{Chris \textsc{Pearson}\altaffilmark{*}\altaffilmark{1}\altaffilmark{,2}\altaffilmark{,3}%
}
\author{Laia \textsc{Barrufet}\altaffilmark{1}\altaffilmark{,2}\altaffilmark{,11}}
\author{Maria del Carmen  \textsc{Campos Varillas}\altaffilmark{1}}
\author{Stephen \textsc{Serjeant}\altaffilmark{2}}
\author{David L \textsc{Clements}\altaffilmark{4}}
\author{Tomotsugu \textsc{Goto}\altaffilmark{5}}
\author{Myungshin \textsc{Im}\altaffilmark{6}}
\author{Woong-Seob \textsc{Jeong}\altaffilmark{7}\altaffilmark{,8}}
\author{Seong Jin \textsc{Kim}\altaffilmark{5}}
\author{Hideo \textsc{Matsuhara}\altaffilmark{9}}
\author{Chris \textsc{Sedgwick}\altaffilmark{2}}
\author{Ivan \textsc{Valtchanov}\altaffilmark{10}}

\altaffiltext{1}{RAL Space, STFC Rutherford Appleton Laboratory, Didcot, Oxfordshire OX11 0QX, UK}
\altaffiltext{2}{The Open University, Milton Keynes, MK7 6AA, UK}
\altaffiltext{3}{Oxford Astrophysics, University of Oxford, Keble Rd, Oxford OX1 3RH, UK}
\altaffiltext{4}{Astrophysics Group, Imperial College London, Blackett Laboratory, Prince Consort Road, London SW7 2AZ, UK}
\altaffiltext{5}{National Tsing Hua University No. 101, Section 2, Kuang-Fu Road, Hsinchu, Taiwan 30013}
\altaffiltext{6}{CEOU/Astronomy Program, Dept. of Physics \& Astronomy, Seoul National University, 1 Gwanak-rho, Gwanak-gu, Seoul, Korea}
\altaffiltext{7}{Korea Astronomy and Space Science Institute (KASI), 776 Daedeok-daero, Yuseong-gu, Daejeon 34055, Korea}
\altaffiltext{8}{Korea University of Science and Technology, 217 Gajeong-ro, Yuseong-gu, Daejeon 34113, Korea}
\altaffiltext{9}{Institute of Space and Astronautical Science, Japan Aerospace Exploration Agency, Sagamihara, Kanagawa 229-8510, Japan}
\altaffiltext{10}{Telespazio Vega UK for ESA, European Space Astronomy Centre, Operations Department, 28691 Villanueva de la Ca\~nada, Spain}
\altaffiltext{11}{European Space Astronomy Centre, 28691 Villanueva de la Ca\~nada, Spain}


\email{chris.pearson@stfc.ac.uk}



\maketitle

\begin{abstract}
A detailed analysis of {\it Herschel}-PACS observations at the North Ecliptic Pole is presented. High quality maps, covering an area of 0.44 square degrees, are produced and then used to derive potential candidate source lists. A rigorous quality control pipeline has been used to create final legacy catalogues in the PACS Green 100$\, \mu$m and Red 160$\, \mu$m bands, containing 1384 and 630 sources respectively. These catalogues reach to more than twice the depth of the current archival {\it Herschel}/PACS Point Source Catalogue, detecting 400 and 270 more sources in the short and long wavelength bands respectively. Galaxy source counts are constructed that extend down to flux densities of 6mJy and 19mJy (50\% completeness) in the Green 100$\, \mu$m and Red 160$\, \mu$m bands respectively. These source counts are consistent with previously published PACS number counts in other fields across the sky. The source counts are then compared with a galaxy evolution model identifying a population of luminous infrared galaxies as responsible for the bulk of the galaxy evolution over the flux range (5-100mJy) spanned by the observed counts, contributing approximate fractions of 50\% and 60\% to the cosmic infrared background (CIRB) at 100$\, \mu$m and 160$\, \mu$m respectively. 
\end{abstract}

\section{Introduction}
Multiwavelength surveys are crucial in order to fully understand the evolution of galaxies with cosmic time. Moreover, given that at least half the radiative energy  from star formation over the history of the Universe has been absorbed by dust and re-radiated at infrared wavelengths, creating a significant  cosmic infrared background (CIRB), surveys at infrared wavelengths are a necessity. Such multi-wavelength treasure chests on the sky enjoy a vast amount of coverage across the entire electromagnetic spectrum. The North Ecliptic Pole (NEP) is one such field, initially targeted at mid-infrared wavelengths as the legacy field for galaxy evolution and cosmology for the  {\it AKARI} satellite  \citep{murakami07, matsuhara06}. The {\it AKARI} NEP field consist of a pair of overlapping surveys covering a deep region of 0.54 deg.$^{2}$, centred at RA=17$\rm ^h$56$\rm ^m$, Dec.=66$^{\circ}$37$\arcmin$ (NEP-Deep, \cite{wada08}) and a shallower surrounding region covering 5.8 deg.$^{2}$  centred on the NEP itself (NEP-Wide, \cite{lee09}). The NEP region provides a continuous viewing zone for most space telescopes, enjoying natural high coverage for a variety of all sky survey missions (e.g. {\it IRAS}, {\it WISE}, {\it Planck}, {\it eRosita}) and is a possible deep field candidate  for future missions such as {\it Euclid}, {\it JWST} and {\it SPICA} \citep{serjeant12}.

The NEP field has also been observed by ESA's {\it Herschel} Space Observatory \citep{pilbratt10}, with the SPIRE instrument \citep{griffin10}  at 250, 350 and 500$\, \mu$m. The SPIRE observations are described in \citet{pearson17}, Pearson et al. (2018), in preparation). In this work we report on observations made with the  {\it Herschel}-PACS instrument \citep{poglitsch10} at 100$\, \mu$m and 160$\, \mu$m. PACS observations are crucial in constraining the short wavelength side of the dust emission hump in star-forming galaxies, helping to break the temperature redshift degeneracy associated with SPIRE colours (e.g. \cite{schulz10}). Here we report on a detailed and careful re-reduction of the PACS observations in the NEP, in order to produce high quality maps, superior to the current archival products in the  {\it Herschel} Science Archive (HSA). These maps are used to produce a list of candidate sources in both PACS bands. A dedicated quality control and cleaning pipeline is then used to produce final robust legacy galaxy catalogues.

In Section~\ref{PACSanalysis} the original observations are introduced, and our data reduction and analysis are explained. In Section~\ref{sourceextraction}, the source extraction and subsequent quality control pipeline is introduced. In Section~\ref{results}, we present our results, comparing our catalogues with a subset of the  archival {\it Herschel}/PACS Point Source Catalogue (HPPSC, \cite{marton17}). We produce new galaxy source counts in the NEP region at 100 and 160$\, \mu$m, comparing them with PACS observations in other cosmology legacy fields. We then use a contemporary galaxy evolution model to interpret the source counts, discussing the relative contribution of the dusty galaxy populations. A summary is given in Section~\ref{summary}. Throughout this work a concordance cosmology of $H_0=70$\,km\,s$^{-1}$\,Mpc$^{-1}$, $\Omega_{\rm M}=0.3$ and $\Omega_\Lambda=0.7$ is assumed.

\newpage

\section{Observations and Data Analysis}  \label{PACSanalysis}

\subsection{Observations}
\begin{figure}
 \begin{center}
  \includegraphics[width=8cm]{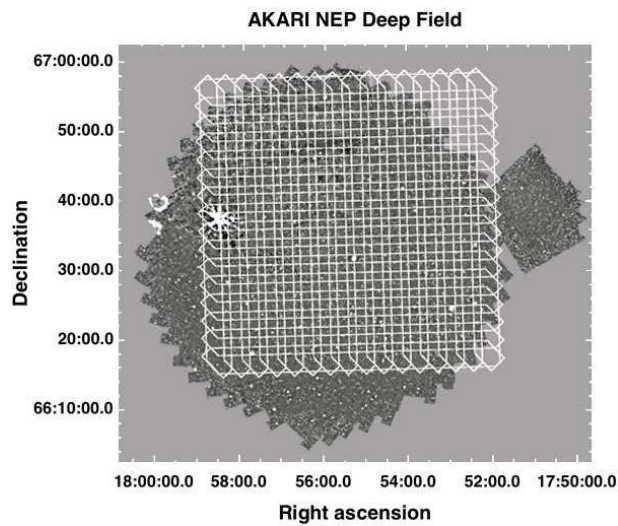} 
 \end{center}
\caption{PACS corverage of the NEP region. Lines are the PACS AORs, covering approximately 0.44 square degrees, overlaid on the {\it AKARI} 15$\, \mu$m image of the NEP(-Deep) region.}
\label{fig:NEPaor}
\end{figure}

The NEP region was observed with {\it Herschel}  as an Open Time 2 programme (OT2\_sserje01\_2) with both the SPIRE and  PACS instruments. The data reduction and analysis for the SPIRE observations is reported in Pearson et al. 2018, in preparation.
The PACS observations were made as a set of 10 nominal and orthogonal direction scan-map AOR (Astronomy Observation Request) pairs of 40 arcmin scan leg length and 20$\arcsec$s$^{-1}$ scan speed, as summarised in Table~\ref{tab:aor}. Observations were made simultaneously in the Green (100$\, \mu$m) and Red (160$\, \mu$m) bands. No observations were made in the Blue (70$\, \mu$m) band. Each individual AOR contained four repetitions of the map, resulting in estimated HSpot  3$\sigma$ instrumental noise sensitivities of 4.59 mJy and 8.73 mJy at 100$\, \mu$m and 160$\, \mu$m respectively. The PACS observations were taken on  consecutive {\it Herschel} Operational Days, 1440-1443 (22-26 April 2013, only 3 days before {\it Herschel} 's "End of Life" on 29th April 2013). Figure~\ref{fig:NEPaor} shows the PACS AORs overlaid on the {\it AKARI} NEP survey image at 15$\, \mu$m. The PACS observations cover approximately 0.44 square degrees overlapping with the majority of the {\it AKARI} NEP-Deep survey.

\begin{table*}
  \tbl{PACS AORs giving the observation ID (obsid), positional information (J2000), Operational Day, AOR, date of observation and duration in secs.}
  {%
  \begin{tabular}{lllllll}
      \hline
obsid	&	RA	&	Dec	&	OD	&	AOR	&	Date	&	Duration (s)	\\      \hline
1342270697	&	17$^h$55$^m$23.95$^s$	&	+66d37$\arcmin$33.15$\arcsec$	&	1440	&	pacs$\_$nep$\_$1$\_$2$\_$o	&	22/04/2013	&	9018	\\
1342270698	&	17$^h$55$^m$24.07$^s$	&	+66d37$\arcmin$31.82$\arcsec$	&	1440	&	pacs$\_$nep$\_$1$\_$1$\_$n	&	22/04/2013	&	9018	\\
1342270699	&	17$^h$55$^m$23.91$^s$	&	+66d37$\arcmin$32.11$\arcsec$		&	1440	&	pacs$\_$nep$\_$2$\_$2$\_$o	&	22/04/2013	&	9018	\\
1342270700	&	17$^h$55$^m$24.00$^s$	&	+66d37$\arcmin$31.91$\arcsec$	&	1440	&	pacs$\_$nep$\_$2$\_$1$\_$n	&	23/04/2013	&	9018	\\
1342270701	&	17$^h$55$^m$23.95$^s$	&	+66d37$\arcmin$32.92$\arcsec$	&	1440	&	pacs$\_$nep$\_$3$\_$2$\_$o	&	23/04/2013	&	9018	\\
1342270702	&	17$^h$55$^m$23.96$^s$	&	+66d37$\arcmin$31.96$\arcsec$	&	1440	&	pacs$\_$nep$\_$3$\_$1$\_$n	&	23/04/2013	&	9018	\\
1342270703	&	17$^h$55$^m$23.94$^s$	&	+66d37$\arcmin$32.54$\arcsec$	&	1440	&	pacs$\_$nep$\_$4$\_$2$\_$o	&	23/04/2013	&	9018	\\
1342270704	&	17$^h$55$^m$23.93$^s$	&	+66d37$\arcmin$32.01$\arcsec$	&	1440	&	pacs$\_$nep$\_$4$\_$1$\_$n	&	23/04/2013	&	9018	\\
1342270866	&	17$^h$55$^m$23.94$^s$	&	+66d37$\arcmin$32.71$\arcsec$	&	1441	&	pacs$\_$nep$\_$8$\_$2$\_$o	&	24/04/2013	&	9018	\\
1342270867	&	17$^h$55$^m$24.04$^s$	&	+66d37$\arcmin$31.86$\arcsec$	&	1441	&	pacs$\_$nep$\_$8$\_$1$\_$n	&	24/04/2013	&	9018	\\
1342270746	&	17$^h$55$^m$23.92$^s$	&	+66d37$\arcmin$32.18$\arcsec$	&	1442	&	pacs$\_$nep$\_$9$\_$2$\_$o	&	24/04/2013	&	9018	\\
1342270747	&	17$^h$55$^m$23.99$^s$	&	+66d37$\arcmin$36.50$\arcsec$	&	1442	&	pacs$\_$nep$\_$9$\_$1$\_$n	&	24/04/2013	&	9018	\\
1342270748	&	17$^h$55$^m$23.96$^s$	&	+66d37$\arcmin$32.60$\arcsec$	&	1442	&	pacs$\_$nep$\_$10$\_$2$\_$o	&	24/04/2013	&	7093	\\
1342270749	&	17$^h$55$^m$24.06$^s$	&	+66d37$\arcmin$31.86$\arcsec$	&	1442	&	pacs$\_$nep$\_$10$\_$1$\_$n	&	25/04/2013	&	7093	\\
1342270902	&	17$^h$55$^m$23.91$^s$	&	+66d37$\arcmin$32.10$\arcsec$	&	1443	&	pacs$\_$nep$\_$5$\_$2$\_$o	&	25/04/2013	&	9018	\\
1342270903	&	17$^h$55$^m$23.99$^s$	&	+66d37$\arcmin$31.94$\arcsec$	&	1443	&	pacs$\_$nep$\_$5$\_$1$\_$n	&	25/04/2013	&	9018	\\
1342270904	&	17$^h$55$^m$23.92$^s$	&	+66d37$\arcmin$32.33$\arcsec$	&	1443	&	pacs$\_$nep$\_$6$\_$2$\_$o	&	25/04/2013	&	9018	\\
1342270905	&	17$^h$55$^m$23.98$^s$	&	+66d37$\arcmin$31.95$\arcsec$	&	1443	&	pacs$\_$nep$\_$6$\_$1$\_$n	&	26/04/2013	&	9018	\\
1342270906	&	17$^h$55$^m$23.93$^s$	&	+66d37$\arcmin$32.78$\arcsec$	&	1443	&	pacs$\_$nep$\_$7$\_$2$\_$o	&	26/04/2013	&	9018	\\
\hline
    \end{tabular}}
    \label{tab:aor}
\begin{tabnote}
\end{tabnote}
\end{table*}

\subsection{Data Reduction}

The {\it Herschel} Science Archive (HSA) was searched for the PACS observations at the NEP. The HSA contains final processed archival co-added maps of all available observations provided by the instrument teams. PACS mapping data  are available as 3 separate science products: the Level 2 maps,  created from individual scanning observations in either the nominal or orthogonal direction;  the Level 2.5 products consisting of combined nominal/orthogonal cross linked scans; Level 3 products consisting of stacked maps for all observations for a given programme at the same position on the sky. The PACS Instrument Control Centre (ICC) recommend the Level 3 products for the best signal-to-noise. In addition, the PACS ICC provides maps created by a variety of map making pipelines. The standard pipeline High Pass Filtered maps are available for the Level 2 and Level 2.5 products. However, it has been shown by \citet{popesso12}, that the high pass filtering can cause a loss in flux density for faint sources of as much as 20-30\% as well as propagating zeros for the associated error maps. Additional map products at the Level 2.5 and Level 3 stages are provided by the JScanam \citep{roussel13}  and the Unimap  \citep{piazzo15} algorithms.

\begin{figure*}
 \begin{center}
  \includegraphics[width=5.5cm]{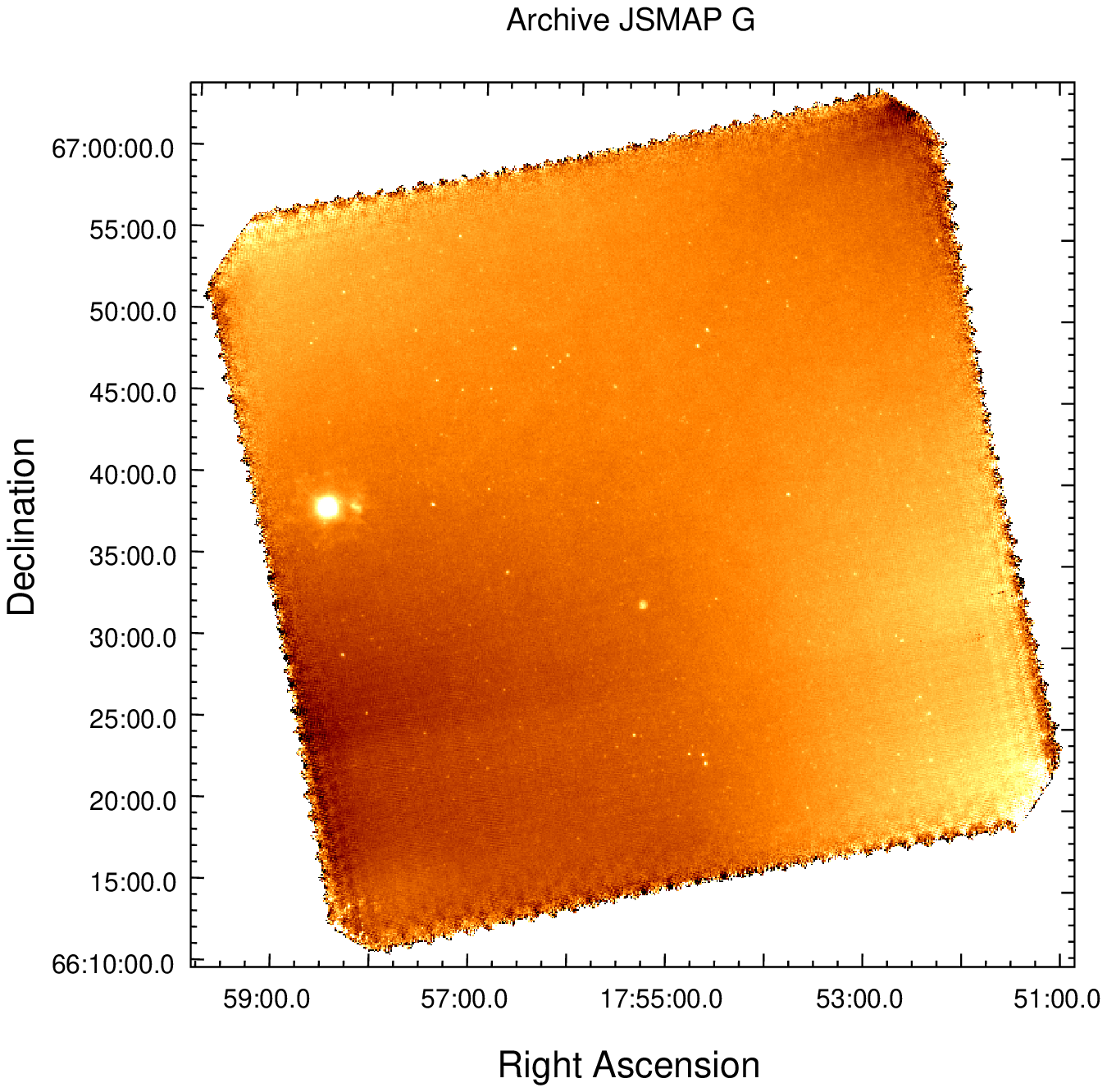} 
  \includegraphics[width=5.5cm]{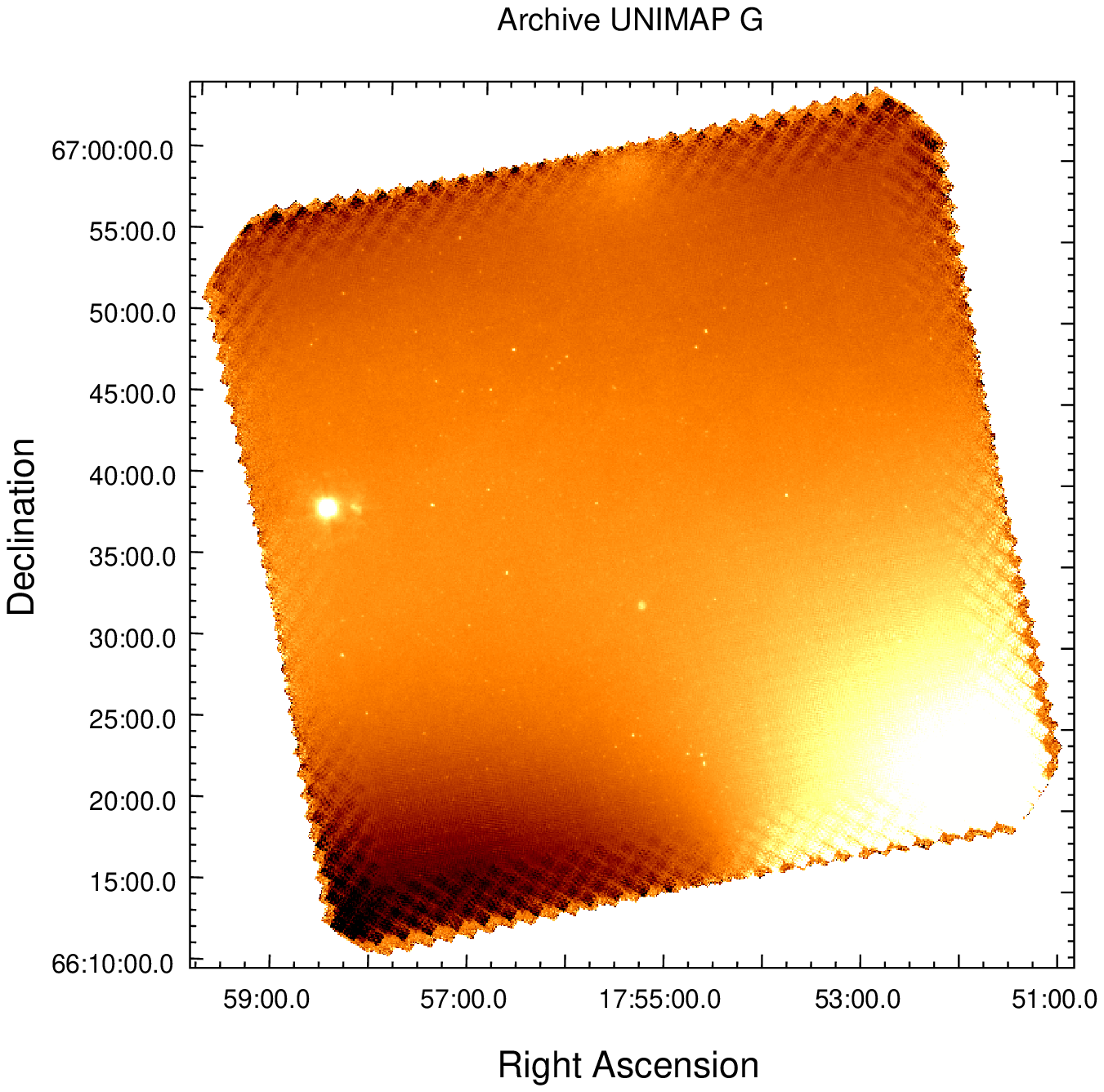} 
  \includegraphics[width=5.5cm]{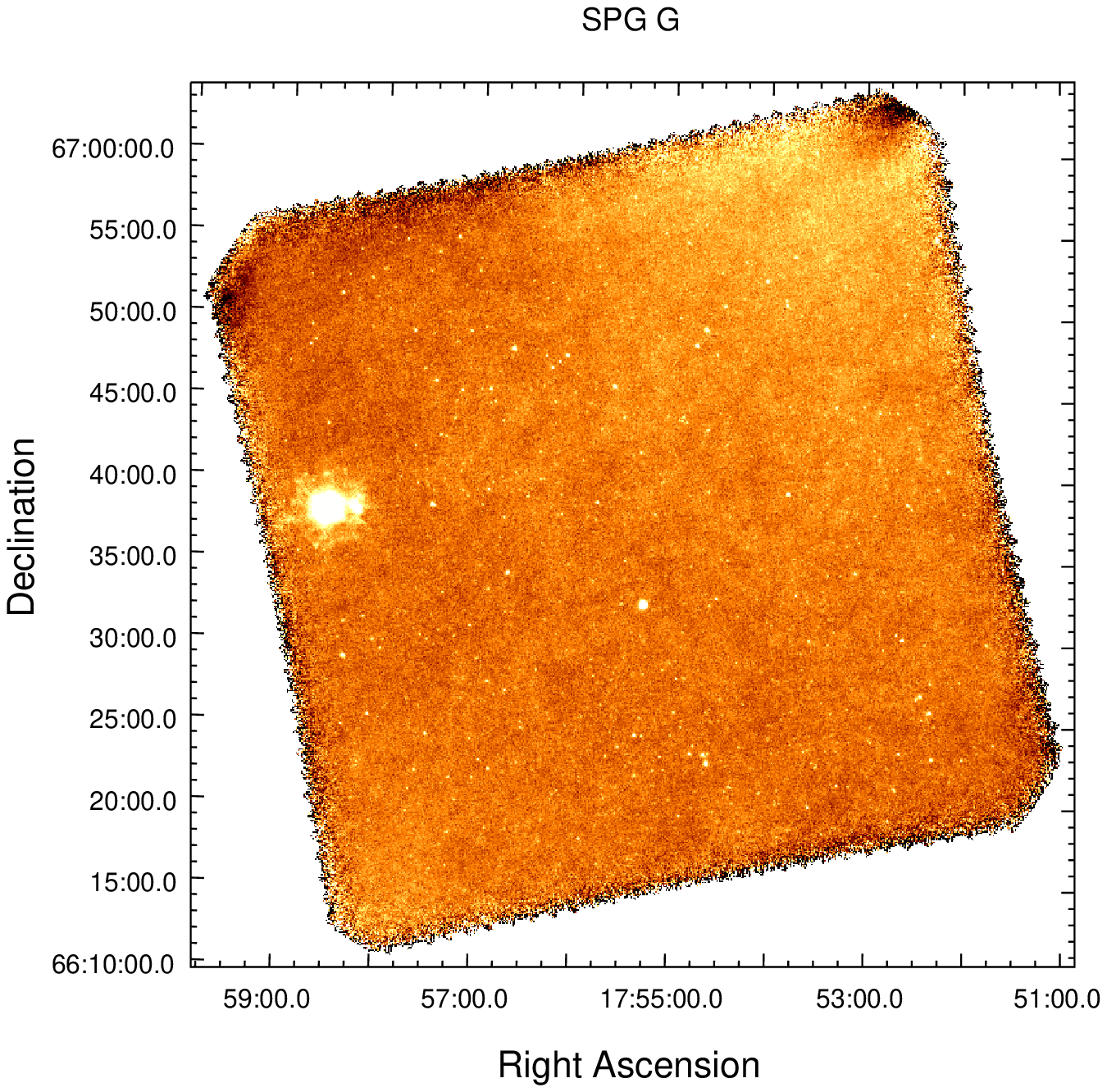} 
 \end{center}
  \begin{center}
  \includegraphics[width=5.5cm]{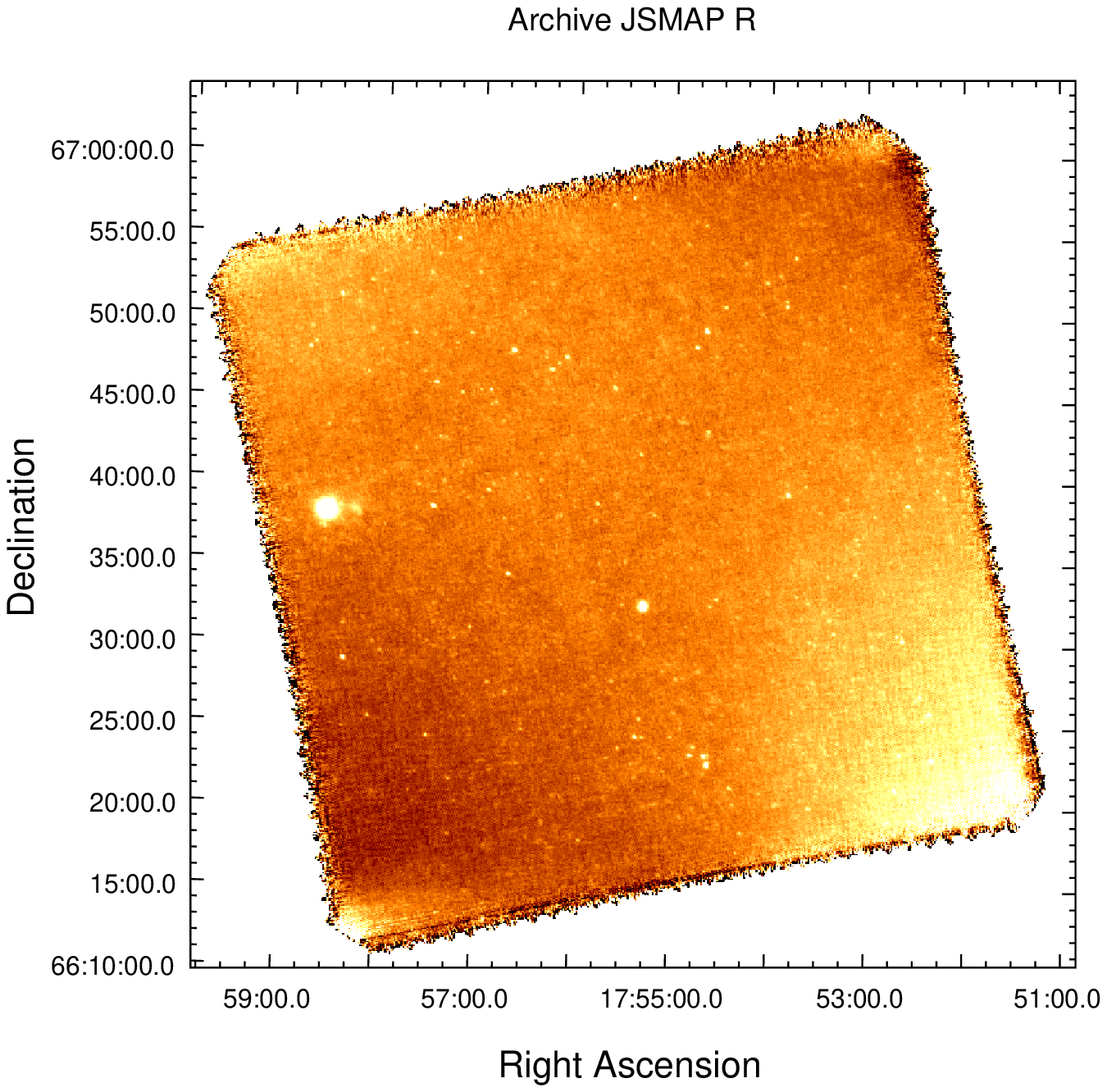} 
  \includegraphics[width=5.5cm]{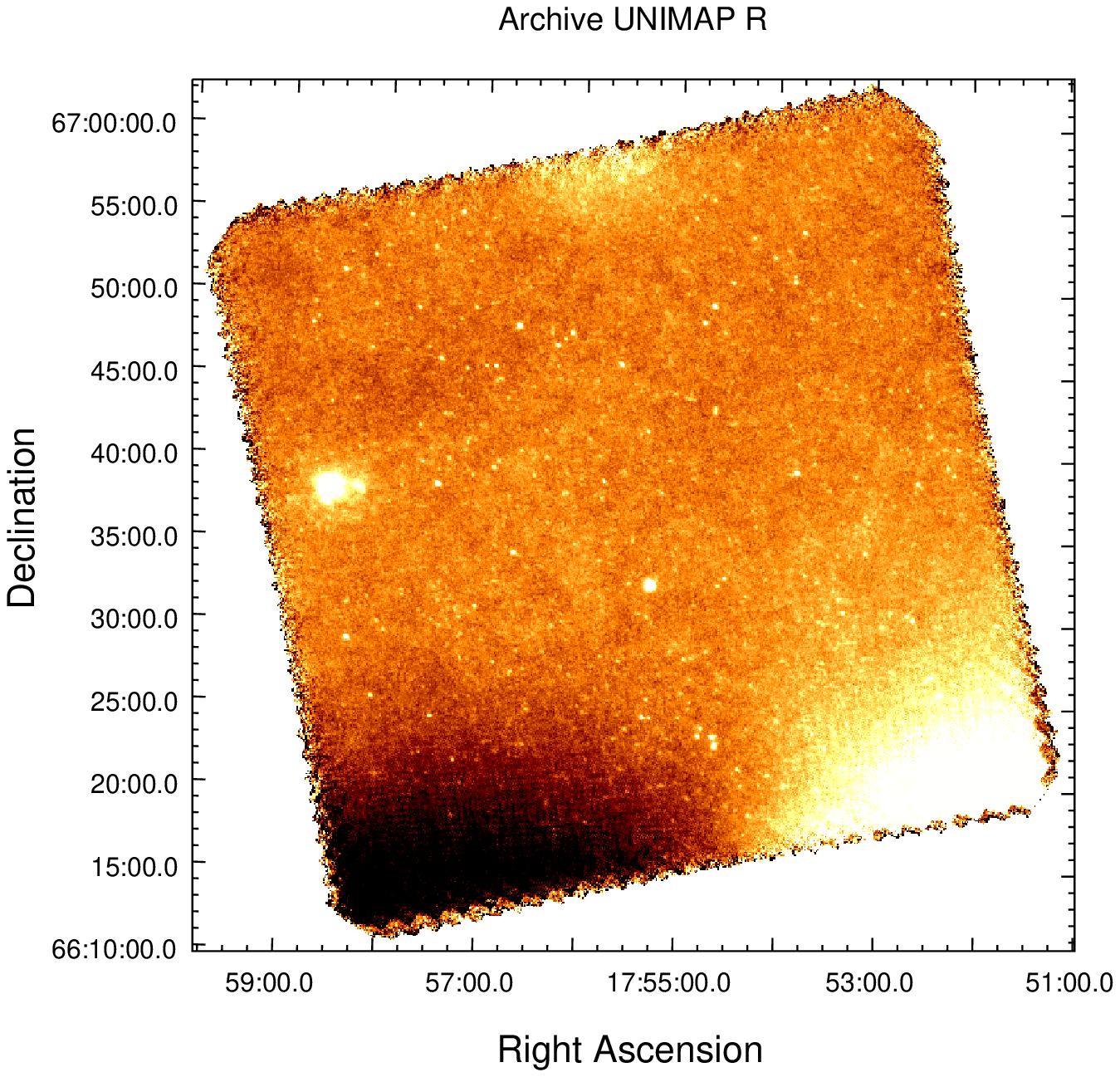} 
  \includegraphics[width=5.5cm]{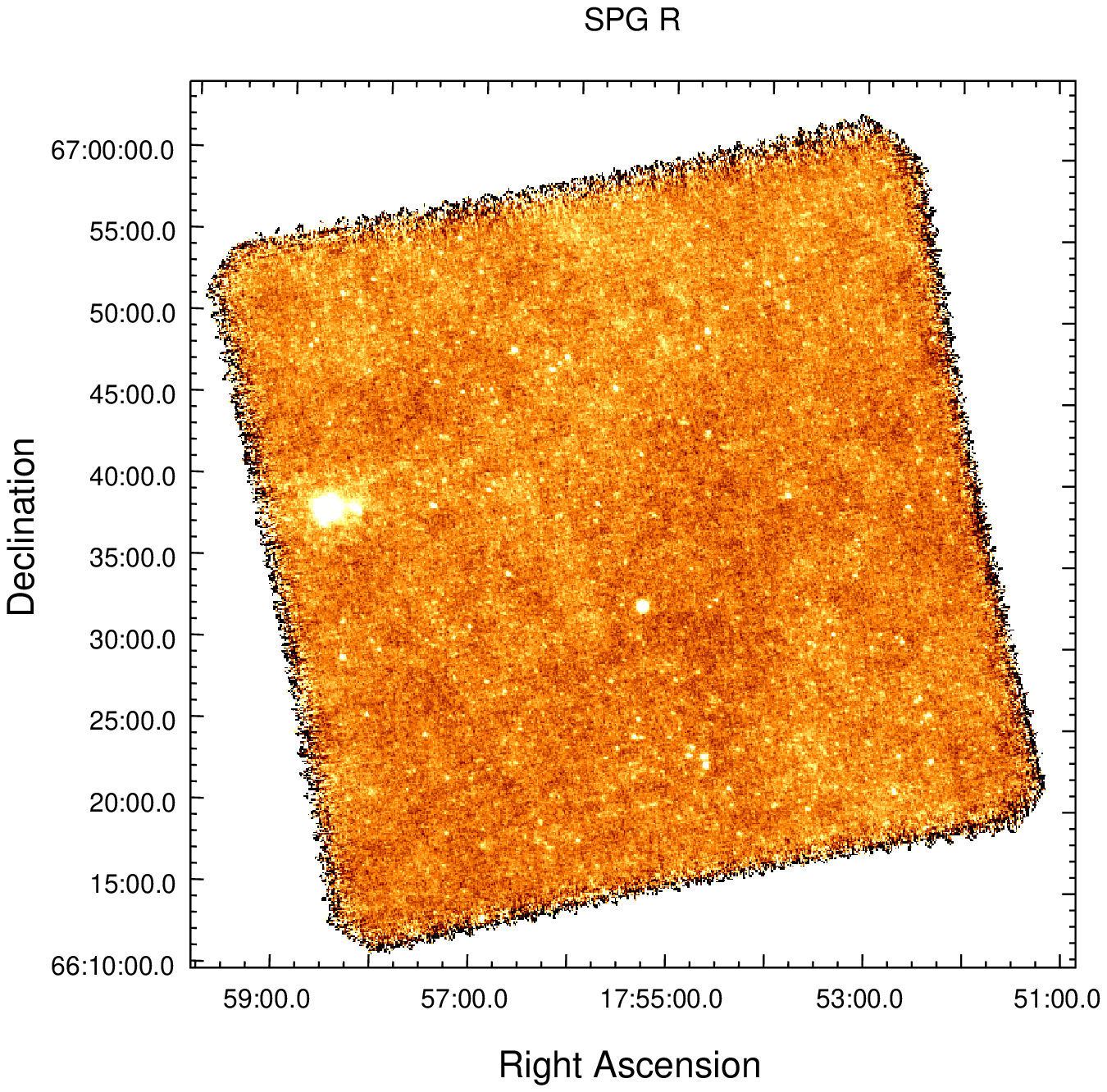} 
 \end{center}
\caption{A comparison of the PACS maps publicly available from the Herschel Science Archive, with the maps produced in this work. Top-row shows archival JScanam and Unimap maps respectively, compared with our pipeline generated maps for the PACS Green 100$\, \mu$m band. Bottom-row shows archival JScanam and Unimap maps respectively, compared with our pipeline generated maps for the PACS Red 160$\, \mu$m band.}
\label{fig:PACSmaps}
\end{figure*}

Unfortunately, both the JScanam and Unimap Level 3 PACS archive maps for our NEP observations, shown in Figure \ref{fig:PACSmaps}, are of poor quality and scientifically unusable. Note that from Operational Day 1375 onwards, half of the red channel array was lost, and masked out automatically in the pipeline processing, however, this is unrelated to our issue since the archival maps for both the red and green PACS channels were affected. 

Therefore, we have examined the individual archive entries in order to identify and reject any anomalous observations responsible for the final poor map products. We find that a pair of (nominal/orthogonal) observations on Operational Day 1442 (obsids 1342270748, 1342270749) exhibit very poor quality in the Level 2.5 products. Looking deeper into the Level 2 maps we find bad quality maps for obsid 1342270749 in particular. There are no quality flags raised on the archival data, nor were there any cooler-recycle operations on this day, however we note that the following unrelated PACS observations (1342270750,1342270751) are listed as $^\prime$FAILED$^\prime$. In addition to the problems found above, we also find striping in the Unimap green band observations for obsid 1342270906. 

To produce science quality maps for this work, we therefore reprocessed the PACS data to Level 3 using the Herschel Common Science System,  {\it Herschel Interactive Processing Environment} (HIPE, \cite{ott11}, \cite{ott13}). The data were re-processed with HIPE version 14.2, omitting the problematic observations, with the Standard Product Generation (SPG) pipeline to Level 3, using the  PACS Calibration Tree version 77. Final Level 2.5 and Level 3 maps (with over-sampling flag set to False), were produced for both the JScanam and Unimap algorithms. From visual inspection of the maps and considerations following from the source extraction (see Section~\ref{sourceextraction}), the final selected products were the Level 3 JScanam maps for both the PACS green and red bands. The final JScanam maps are shown in the right-most panels of Figure~\ref{fig:PACSmaps}, highlighting the improved quality of the maps compared to the original archive products.

\section{Source Extraction and Photometry} \label{sourceextraction}

\subsection{Source Extraction}  \label{extraction}

Sources were detected and extracted from the Level 3 maps using the HIPE SUSSEXtractor task (\cite{savage07}, \cite{pearson14}). In order to avoid detection of excessively large numbers of spurious sources, due to noisy edge effects and map artefacts (see right-most panels of Figure~\ref{fig:PACSmaps}), a mask was applied to the maps covering both bright extended sources, e.g. NGC 6543, the Cat's Eye Nebula and the map edges. The  mask extends approximately 60 arcsec in from each edge around the maps, corresponding to around 40 and 20 pixels for the Green 100$\, \mu$m and Red 160$\, \mu$m band maps respectively.
SUSSEXtractor models both the source and the empty sky using a Bayesian approach to determine whether a source is present at any given position in the map. The initial image is smoothed by a point response function (PRF) with a point source candidate identified in any pixel with some local maxima, above some detection threshold parameter, compared to its neighbours, within some defined pixel region. The peak intensity within the smoothed image at the position of the point source is taken as the estimate of the flux density. Various combinations of the SUSSEXtractor parameter set described above were explored, with the final adopted parameters listed below:
\begin{itemize}
\item {\tt detThreshold} = 3 : The threshold at which a local maximum is detected.
\item {\tt pixelRegion} = 3 : Radius around each pixel to consider when searching for local maxima.
\item {\tt signaltoNoise} = True : Use conventional S/N for detection, rather than the log(Bayes factor).
\item {\tt fitBackground} = True : Fits the background as a free parameter.
\item {\tt roi} = True : Sets a region of interest defined by the applied edge mask.
\end{itemize}

In addition, SUSSEXtractor requires a Full-Width-Half-Maximum (FWHM) for each waveband. We adopt the mean values, appropriate for the observation's scan speed of 20$\arcsec$s$^{-1}$ from the PACS Observers Manual \footnote{PACS Observers Manual: \\ \tt{http://herschel.esac.esa.int/Docs/PACS/html/pacs\_om.html}} of 6.8$\arcsec$ and 11.3$\arcsec$ for the PACS Green 100$\, \mu$m and Red 160$\, \mu$m bands respectively. Note that slightly differing the FWHM values does not change the number of sources extracted from the map. Unless otherwise supplied, SUSSEXtractor assumes a 5$\times$5 pixel Gaussian PSF. However, much work has been invested in characterising the PACS PSF, and we adopt the more realistic PSFs from the ESA-PACS document archive (Technical Report PICC-ME-TN-033\footnote{PACS Photometer PSF (PICC-ME-TN-033): \\ \tt{https://www.cosmos.esa.int/documents/12133/996891/\\PACS+photometer+point+spread+function}}). We use the following instrument PSF files:
\begin{description}
\item[\tt{psf20\_grn\_20\_vesta\_od160\_ama+63\_recentered.fits}] 
\item[\tt{psf20\_red\_20\_vesta\_od160\_ama+63\_recentered.fits}] 
\end{description}
for the PACS Green 100$\, \mu$m and Red 160$\, \mu$m bands respectively, corresponding to observations of Vesta on Operational Day 160 at a scan speed 20$\arcsec$s$^{-1}$ and an AMA of 63 degrees (array-to-map angle i.e. the scan direction in degrees CCW from the spacecraft Z-direction).

Using these parameters on our masked JScanam maps, SUSSEXtractor detects a total of 3472 and 1815 candidate sources in the PACS Green 100$\, \mu$m and Red 160$\, \mu$m bands respectively. Note that extraction using the Unimap products yielded extremely low numbers of sources, only $\sim$150 in the PACS Red 160$\, \mu$m band, and visual inspection of the maps shows that many sources were being missed. Therefore the JScanam maps were used in preference.

\subsection{Source List Cleaning}

In order to produce a final robust and reliable catalogue of sources, our candidate list has been processed through a quality control pipeline that broadly follows the procedure of \citet{marton17}, who have produced an archival {\it Herschel}/PACS Point Source Catalogue (HPPSC) covering the entire sky. The HPPSC is intended as a legacy data product for the {\it Herschel} mission, containing a list of conservatively selected sources from all {\it Herschel}-PACS observations in the  {\it Herschel} Science Archive.

The first step in the quality control pipeline is to use the HIPE $SourceFitting$ task to produce a confirmation of the source position provided by SUSSEXtractor. The $SourceFitting$ task fits a 2-D Gaussian to the data within a selection box. After some consideration, a 15$\arcsec\times$15$\arcsec$ selection box was chosen, with the task parameter settings; $\tt{Elongated= False}$ and $\tt{Slope=False}$, corresponding to  whether the source is elongated (or circular) and whether the background has a slope. However, we found that the results of the source fitting were relatively insensitive to these parameters. The position derived from the source fitting was compared with the result from SUSSEXtractor and a source candidate was rejected if the separation between the two positions was $>$3.6$\arcsec$ or approximately 3 times the {\it Herschel} average pointing error. 

The second step in the quality control pipeline is to use the Gaussian fit from the $SourceFitting$ task to  constrain the FWHM of the candidate source. A candidate is again rejected if the measured FWHM falls outside the range, 0.75$\times$PSF $<$ FWHM$_{Gaussian} <$ 2$\times$PSF. These values are derived from the simulations presented in  \citet{marton17}. These steps, summarised in Table~\ref{tab:sourceselection}, reduce the number of candidate sources from 3742 to 1854 and 1815 to 697 in the PACS Green 100$\, \mu$m and Red 160$\, \mu$m bands respectively.

The final step in the quality control pipeline is a simple flux consistency test using the circular aperture photometry ($\tt{annularSkyAperturePhotometry}$) task within HIPE, in order to compare the candidate source flux with the estimate from the SUSSEXtarctor task. The circular aperture photometry task takes as input a source aperture  and a background annulus in arcsec. This aperture photometry is carried out on all of our remaining candidates at the SUSSEXtractor position with two sets of apertures. For the PACS Green 100$\, \mu$m band, two aperture sets of [4$\arcsec$, 25$\arcsec$, 35$\arcsec$] and [9$\arcsec$, 25$\arcsec$, 35$\arcsec$] are used, where the first value in the parenthesis is the source aperture and the second and third are the inner and outer background annuli respectively. For the PACS  Red 160$\, \mu$m band, two aperture sets of [9$\arcsec$, 25$\arcsec$, 35$\arcsec$] and [14$\arcsec$, 25$\arcsec$, 35$\arcsec$] are used. Following  \citet{marton17}, source candidates are rejected if the ratio of the aperture sets fall outside the range 0.5 $< $aper$_{large}$/aper$_{small}$ $<$  2. Applying this 3rd criterion reduces the candidate list to 1349 in the Green 100$\, \mu$m band and 607 sources in the Red 160$\, \mu$m band.

Finally, the maps were visually inspected within the masked region for any sources that were lost due to the rather conservative mask. Rejecting candidates on the map edges, a total of 35 and 23 'lost' sources were added to produce the final confirmed source lists of 1384 sources in  in the Green 100$\, \mu$m band and 630 sources in the Red 160$\, \mu$m band.

The source cleaning quality control pipeline is summarised in Table~\ref{tab:sourceselection}.

\begin{table*}
  \tbl{Source list cleaning criterea. The original candidate source list from SUSSEXtractor is cleaned via source seperation and FWHM range using 2D Gaussian source fitting. This is followed by a flux measurement, using annular sky aperture photometry. Finally any sources lost to the map mask are added, to finalise the confirmed source list for the PACS Green 100$\, \mu$m and Red 160$\, \mu$m bands.}{%
  \begin{tabular}{llll}
      \hline
Quality Control	&	Criteria	&	 \multicolumn{2}{c}{Number of candidates}	\\      
	                 &		        &	Green (100$\, \mu$m)	&	Red (160$\, \mu$m)		\\      \hline
SUSSEXtractor	&	none 	&	3472 	&	1815	\\   
Source Seperation	&	$<$3.6$\arcsec$ 	&	3235 	&	1649	\\   
FWHM	&	0.75$\times$PSF $<$ FWHM$_{Gaussian} <$ 2$\times$PSF 	&	1854 	&	697	\\   
Flux agreement 	&	0.5 $<$ Aper$_{large}$/Aper$_{small} <$ 2  	&	1349 	&	607	\\   
Lost Sources	&	none 	&	+35 	&	+23	\\    \hline
  &	Total	 	&	1384 	&	630	\\   
 \hline
     \end{tabular}}
    \label{tab:sourceselection}
\begin{tabnote}
\end{tabnote}
\end{table*}

\subsection{Source List Photometry}

\begin{figure}
 \begin{center}
  \includegraphics[width=8cm]{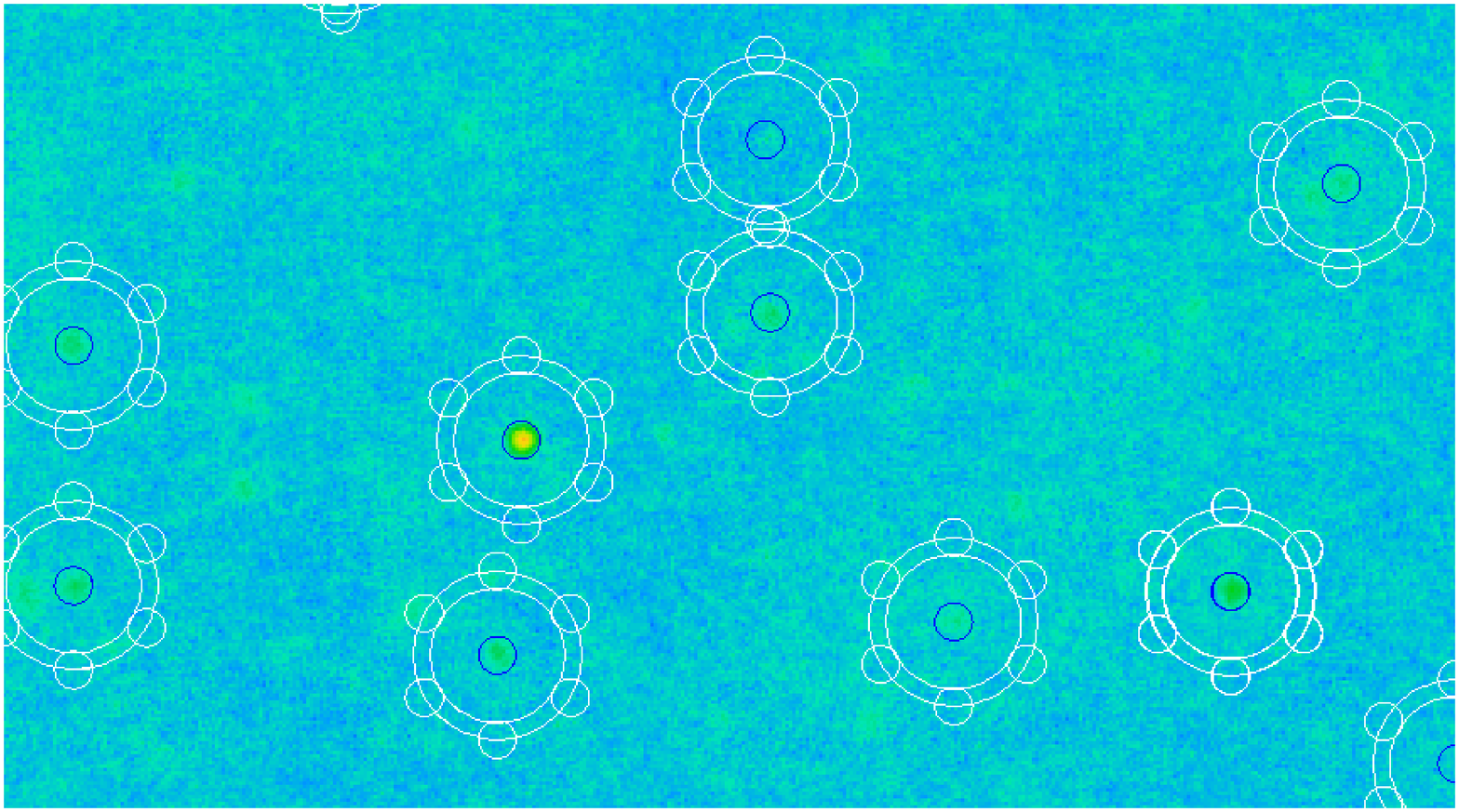} 
 \end{center}
\caption{Example of PACS aperture photometry on the Green 100$\, \mu$m band map. The task lays down a source aperture and  inner/outer background annuli. The task calculates  realistic errors for the aperture photometry by laying down a series of smaller apertures around but offset from the source position.}
\label{fig:aperturephotometry}
\end{figure}

Photometry is carried out on the final source list to provide definitive flux density values for the final catalogue. We use a dedicated processing script within the HIPE environment, developed by the PACS ICC for aperture photometry on an input target list. Source locations are given by our SUSSEXtractor positions. The photometry script requires both an image map and a coverage map, taking as input an aperture diameter (in arcsec) and  inner and outer values for the background annulus. We assume aperture sets in between our test cases of the form [source annulus , inner annulus, outer annulus] equal to  [6$\arcsec$, 15$\arcsec$, 20$\arcsec$] for the  PACS Green 100$\, \mu$m band and [10.5$\arcsec$, 15$\arcsec$, 20$\arcsec$] for the PACS Red 160$\, \mu$m band. The script automatically selects the appropriate aperture correction for the pipeline calibration (PACS FM7 response calibration file), corresponding to 0.595 and 0.673 for the Green 100$\, \mu$m and Red 160$\, \mu$m bands respectively. The script also calculates realistic errors for the aperture photometry by laying down a series of apertures around but offset from the source position, as shown in Figure~\ref{fig:aperturephotometry}. These aperture corrected flux densities and associated photometric errors are used to populate the final catalogue products.

\subsection{Final Catalogues}

The final catalogues contain RA, Dec (J2000 epoch in decimal degrees) positional information, aperture corrected flux densities and associated photometric errors for 1384 sources in the PACS Green 100$\, \mu$m band and 630 sources in the PACS Red 160$\, \mu$m band. The catalogues are publicly available via the CDS VizieR catalogue database \footnote{VizieR: \\ \tt{http://vizier.u-strasbg.fr/viz-bin/VizieR}}. In Tables \ref{tab:catGreen} and \ref{tab:catRed} the first 10 entries for the PACS Green 100$\, \mu$m band and PACS Red 160$\, \mu$m band catalogues are shown respectively.

\begin{table}
  \tbl{PACS Green 100$\, \mu$m band catalogue (first 10 entries). The catalogue contains RA, Dec (J2000 epoch in decimal degrees) positional information, aperture corrected flux densities and associated photometric errors.}
  {%
  \begin{tabular}{lllllll}
      \hline
RA (Deg)	&	Dec (Deg) 	&	Flux (mJy)	&	Flux Error (mJy)		\\      \hline
269.1661916 & 66.8002076 & 245.253 & 2.600 \\
269.3669944 & 66.63897734 & 221.583 & 1.343 \\
269.5892524 & 66.48205997 & 188.625 & 4.168 \\
268.6622189 & 66.38798673 & 182.758 & 1.406 \\
268.4520009 & 66.65580907 & 171.974 & 1.963 \\
268.6552678 & 66.37914993 & 155.342 & 2.476 \\
268.0719311 & 66.91658854 & 152.748 & 2.898 \\
268.6670081 & 66.82278598 & 145.883 & 1.066 \\
268.6912145 & 66.80615083 & 144.062 & 3.197 \\
269.1710021 & 66.57115239 & 123.529 & 2.126 \\
\hline
    \end{tabular}}
    \label{tab:catGreen}
\begin{tabnote}
\end{tabnote}
\end{table}

\begin{table}
  \tbl{PACS Red 160$\, \mu$m band catalogue (first 10 entries). The catalogue contains RA, Dec (J2000 epoch in decimal degrees) positional information, aperture corrected flux densities and associated photometric errors.}
  {%
  \begin{tabular}{lllllll}
      \hline
RA (Deg)	&	Dec (Deg) 	&	Flux (mJy)	&	Flux Error (mJy)		\\      \hline
269.1662109 & 66.8001525 & 278.013 & 8.566 \\
269.3669409 & 66.6390003 & 262.389 & 1.853 \\
268.6669351 & 66.82269989 & 227.373 & 4.143 \\
268.6552791 & 66.37908225 & 224.173 & 3.621 \\
269.5892591 & 66.4820033 & 206.098 & 7.356 \\
268.6619808 & 66.38790302 & 177.853 & 8.777 \\
268.4522128 & 66.65579327 & 171.477 & 8.199 \\
269.1708909 & 66.57115885 & 158.281 & 5.862 \\
268.691138 & 66.80612038 & 144.581 & 5.552 \\
268.838679 & 66.40712293 & 142.19 & 17.278 \\
\hline
    \end{tabular}}
    \label{tab:catRed}
\begin{tabnote}
\end{tabnote}
\end{table}

\newpage
\section{Results}\label{results}

\subsection{Comparison with HPPSC}
\begin{figure*}
 \begin{center}
\includegraphics[width=3.7cm, angle=-90]{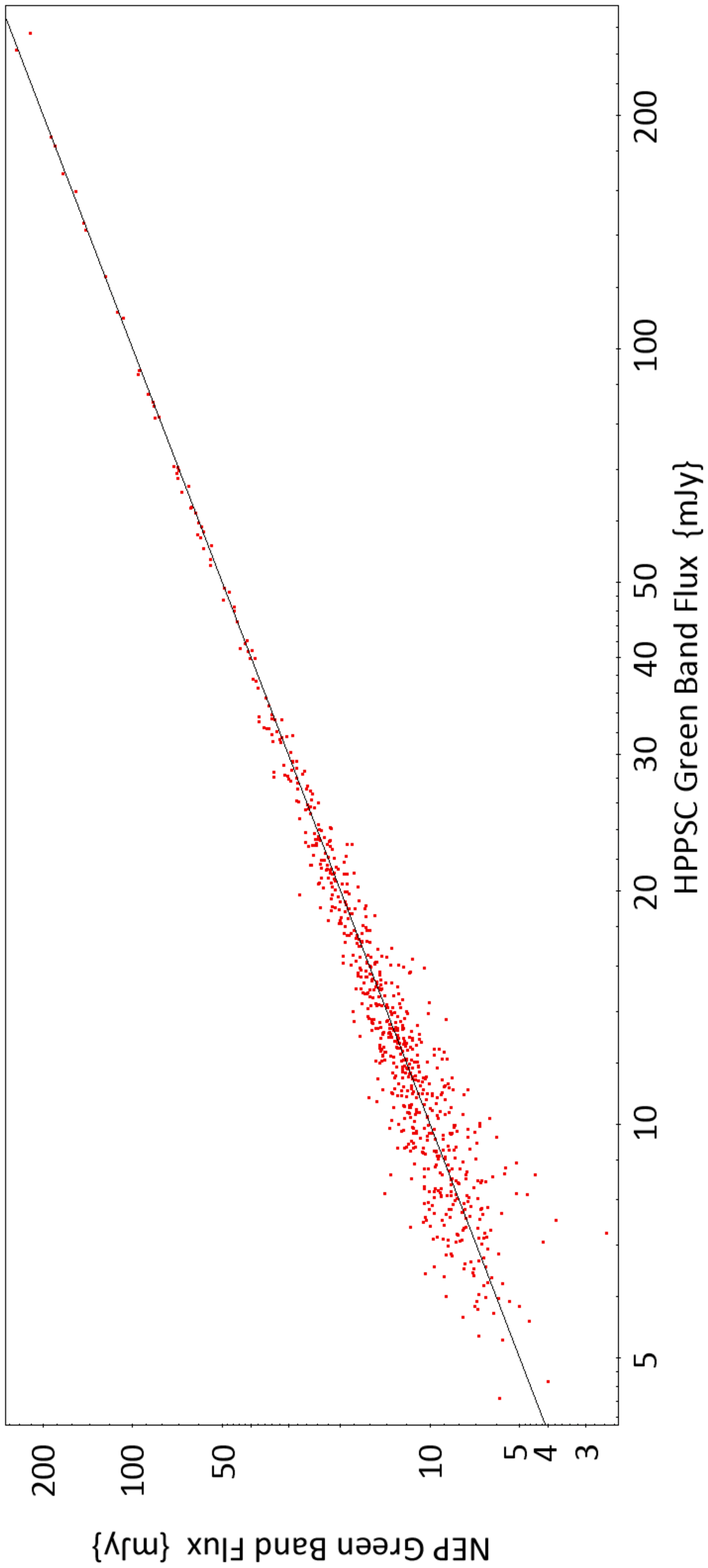} 
\includegraphics[width=3.7cm, angle=-90]{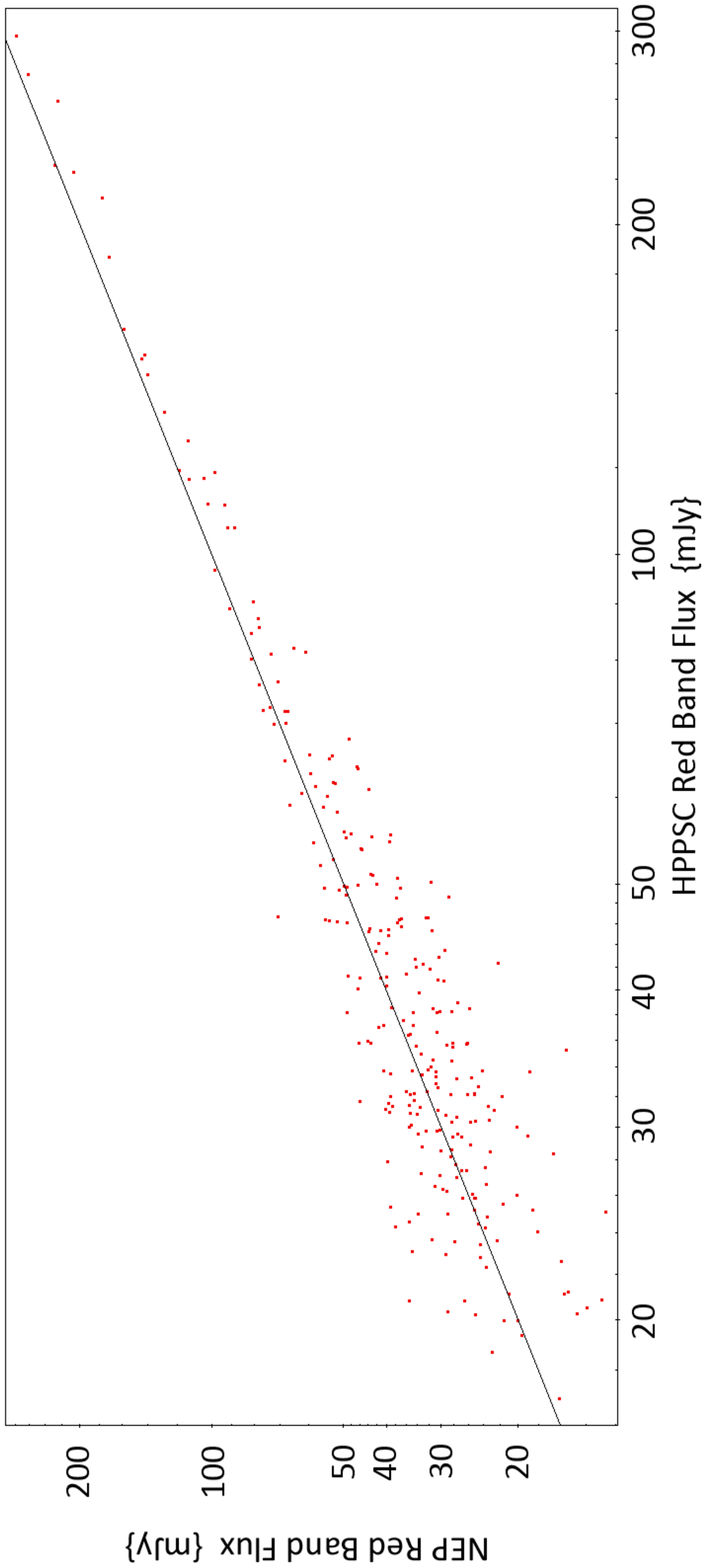} 
 \end{center}
\caption{Comparison of derived source fluxes, in mJy, from our catalogue (vertical axis) compared with the HPPSC (horizontal axis). Left-panel: PACS Green 100$\, \mu$m band. Right-panel PACS Red 160$\, \mu$m band. The plots show a reasonable agreement in the measured fluxes between the two catalogues.}
\label{fig:compareHPPSCflux}
\end{figure*}

We can compare our catalogues in the PACS Green 100$\, \mu$m and Red 160$\, \mu$m bands with the archival  {\it Herschel}/PACS Point Source Catalogue (HPPSC, \cite{marton17}). Selecting sources from the HPPSC coincident with our NEP survey area, we find 984 sources and 360 sources in the PACS Green 100$\, \mu$m and Red 160$\, \mu$m bands respectively. Therefore our new catalogues contain 400 and 270 more sources in the short and long wavelength bands. The new catalogues were cross-matched with the HPPSC in each band, assuming a search radius of 3.6$\arcsec$. This resulted in a match for approximately 80\% of the sources in each band. In Figure~\ref{fig:compareHPPSCflux}, a comparison is made between the measured flux densities from our catalogue and the HPPSC. The fluxes show very good agreement between the Green band catalogues down to 20mJy with a small random scatter at fainter fluxes of the order of $\pm$3 mJy. The Red band is also in reasonable agreement but with a larger scatter of  order  $\pm$5 mJy. This scatter is due to the noise in the map measured in the apertures at this low flux density level and is consistent with the measured Poisson errors from the aperture photometry task.

\begin{figure}
 \begin{center}
\includegraphics[width=3.7cm, angle=-90]{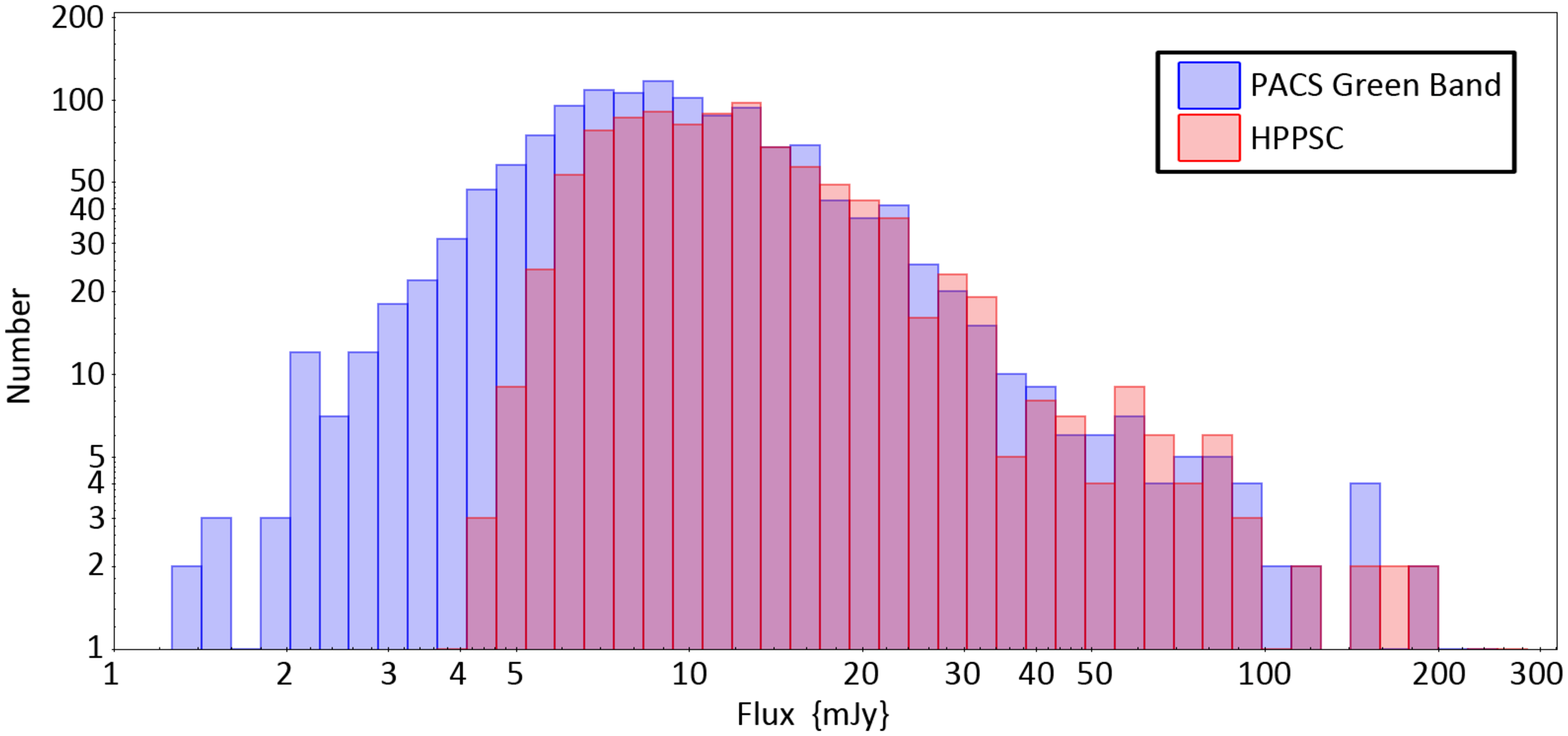} 
\includegraphics[width=3.7cm, angle=-90]{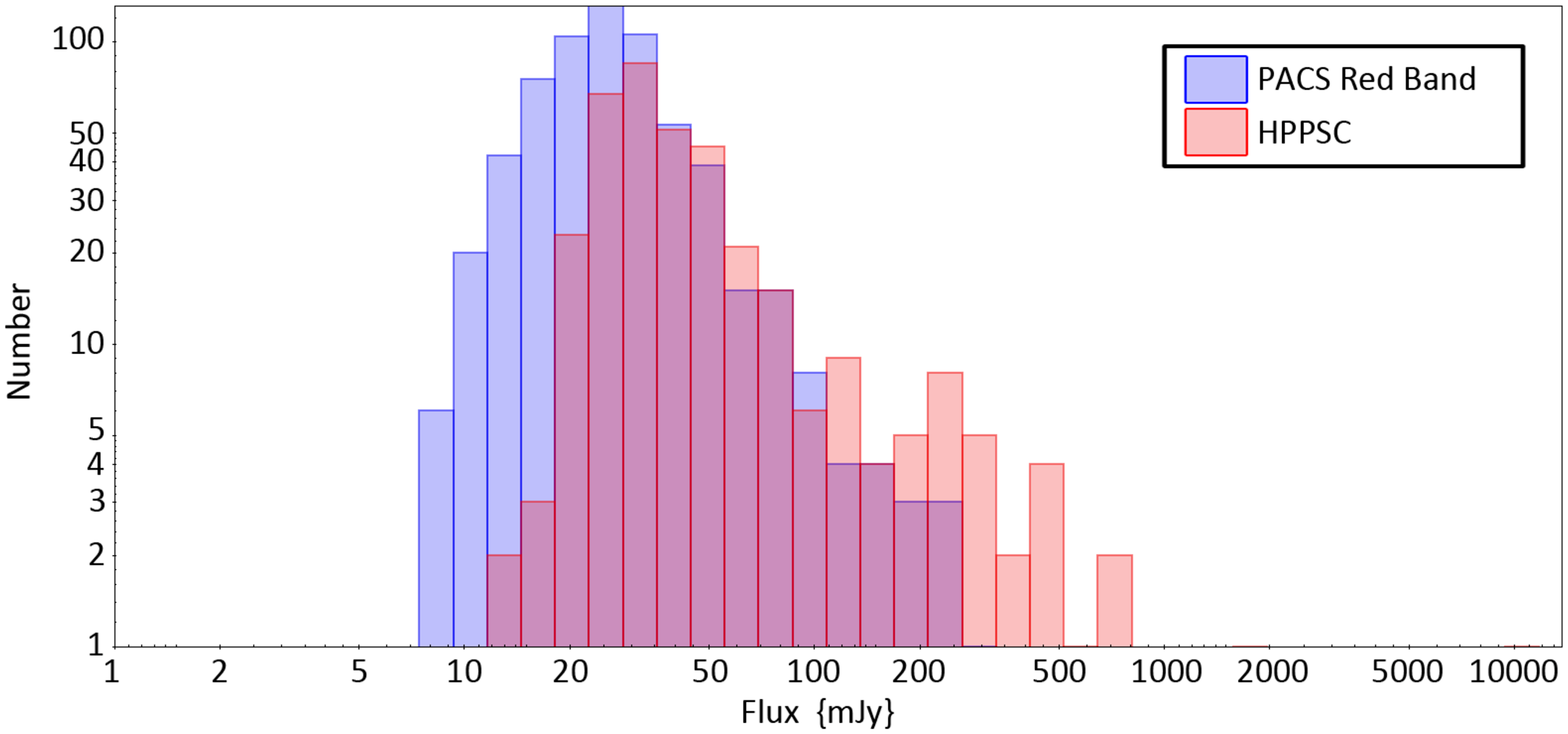} 
 \end{center}
\caption{Comparison of the number of detected sources (vertical axis) from our catalogue (blue bars) compared with the HPPSC (red bars) showing a significant improvement in depth and the number of faint sources detected. Left-panel: PACS Green 100$\, \mu$m band. Right-panel PACS Red 160$\, \mu$m band. The apparent excess in the number of bright sources $>$100mJy in the HPPSC Red 160$\, \mu$m band is due to spurious detections around the bright Cats Eye Nebula as shown in Figure~\ref{fig:compareHPPSCsources}.}
\label{fig:compareHPPSCnumber}
\end{figure}

\begin{figure}
\begin{center}
  \includegraphics[width=8cm]{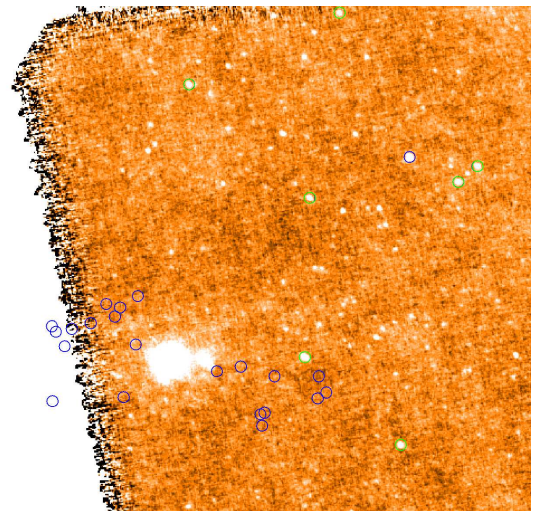} 
 \end{center}
\caption{Overplotting the brightest sources $>$100mJy from our PACS Red 160$\, \mu$m band catalogue (green circles) and the archival HPPSC (blue circles) on the  Red 160$\, \mu$m band map. The enormous bright extended source is the Cats Eye Nebula (NGC 6543) in the NEP field. It can be seen that the HPPSC includes many sources detected as artefacts around the nebula. This may be due to the fact that the HPPSC possibly includes earlier PACS observations centred on the nebula itself.}
\label{fig:compareHPPSCsources}
\end{figure}

 Figure~\ref{fig:compareHPPSCnumber}  shows the number distribution histograms separately for each band, overlaid on the equivalent HPPSC histogram. It can be clearly seen that the new catalogue extends to much fainter fluxes, more than twice as deep, detecting a significant population of fainter sources compared to the HPPSC. Our new catalogues extend down to flux densities of $\sim$1.5mJy and 8mJy in the Green 100$\, \mu$m and Red 160$\, \mu$m bands respectively. Note that these flux densities are still significantly higher than the 5$\sigma$ source confusion limits derived by \citet{magnelli13} in the PEP/GOODS field of 0.75mJy and 3.4mJy at 100$\, \mu$m and 160$\, \mu$m.
From Figure~\ref{fig:compareHPPSCnumber}, it can be seen that at brighter flux densities, our new catalogues appear to be missing sources when compared to the archive HPPSC. This is more significant in the PACS Red 160$\, \mu$m band above 100mJy in the right-panel of Figure~\ref{fig:compareHPPSCnumber}. To investigate these bright sources not included in our catalogues, we have overlaid the brightest sources from the HPPSC catalogue on our image maps and investigated each source independently by eye. For the PACS Green 100$\, \mu$m band, there are 4 sources brighter than 50mJy that are not included in our catalogue. Of these, two appear to be bright artefacts on the map edges, another is a bright extended source (identified as 2MASX J17563982+664800, also observed by {\it IRAS}, extended at 12$\arcsec$at z = 0.089 \cite{ashby96}, \cite{dellavale06}). The final source appears to be another artefact around the Cats Eye Nebula (NGC 6543).  For the PACS Red 160$\, \mu$m band, there are more than 20 sources brighter than 100mJy that are present in the HPPSC but not included in our catalogue. On visual inspection of the PACS Red 160$\, \mu$m band map, almost all of these sources appear to be artefacts caused by the proximity of the Cats Eye Nebula (shown in Figure~\ref{fig:compareHPPSCsources}) -- 61Jy at 100$\, \mu$m, as measured by {\it IRAS} \citep{moshir90} -- or around the map edges. Note that the HPPSC source extraction is made on all maps in the archive and may unintentionally include early PACS observations of the Cats Eye Nebula itself (obsid 1342220097, 1342220098 on Operational Day 723). This explains the apparent discrepancy in the number of bright sources detected between the catalogues.

\subsection{Comparison with {\it AKARI} NEP survey}
The new PACS catalogues have also been cross-correlated with the {\it AKARI} mid-infrared catalogue of \citet{murata13},  over the NEP-Deep region in Figure~\ref{fig:NEPaor}.  Cross-matching with the {\it AKARI}-NEP catalogue using a matching radius of 2$\arcsec$, we find mid-infrared counterparts for almost all the PACS sources down to approximately 10mJy. At flux densities fainter than 10mJy, we find matches with more than 60\% of  the {\it AKARI} sources.

\subsection{Galaxy Number Counts}

\begin{figure*}
 \begin{center}
  \includegraphics[width=8cm]{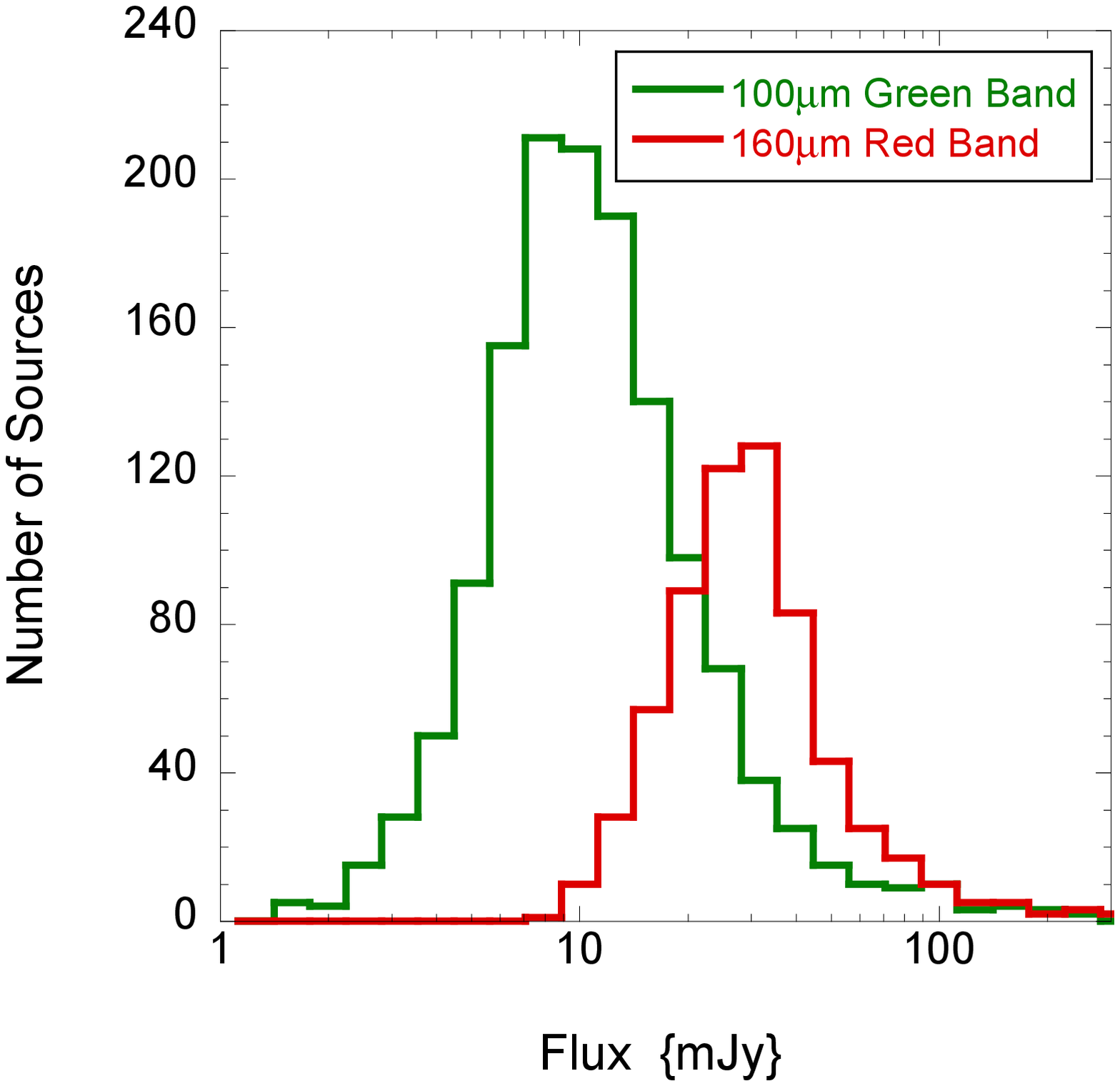} 
  \includegraphics[width=8cm]{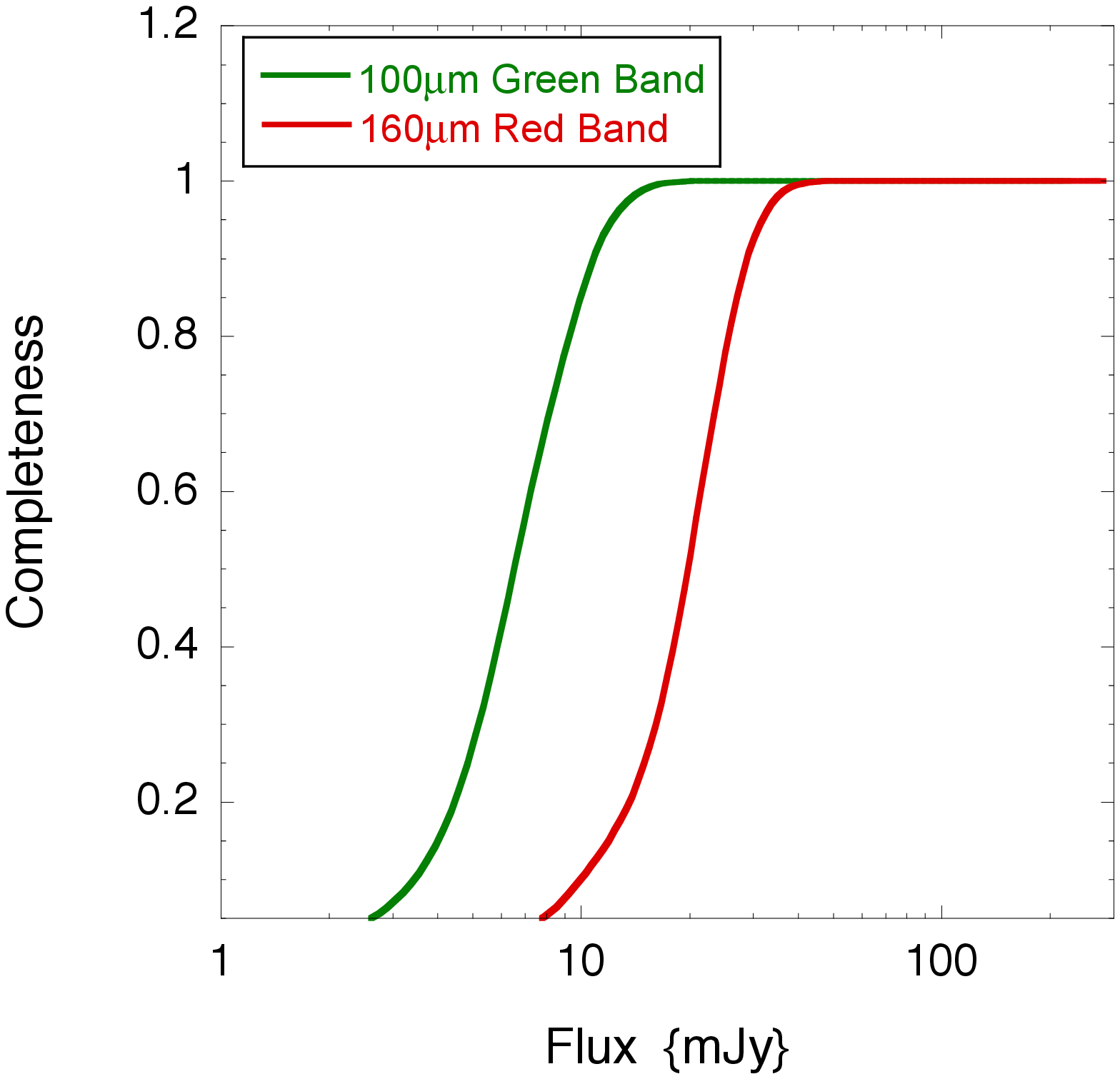} 
 \end{center}
\caption{Left-panel: Number of sources detected over the survey area of 0.44sq.deg. binned in flux steps of log$(flux/mJy)$=0.1. Right-panel shows completeness of the survey as a function of flux density (in mJy). The {\it green} and {\it red} lines represent the PACS Green 100$\, \mu$m and  Red 160$\, \mu$m bands respectively.}
\label{fig:numbercounts}
\end{figure*}

In the left panel of Figure ~\ref{fig:numbercounts}, we plot the number histogram of detected sources over the survey area. The source counts are binned in flux; $\delta$lgS=0.1, for the PACS Green 100$\, \mu$m and  Red 160$\, \mu$m bands. The number distribution for the Green band peaks at 7mJy whilst the Red band peaks significantly brighter at $\sim$25mJy, reflecting the quality of the original maps. 

An estimation of the completeness of the PACS catalogues (fraction of sources missed as a function of flux density) is carried out by performing Monte Carlo simulations.  For each PACS band, artificial sources are injected into the original maps modelled with the same PSFs defined in Section~\ref{extraction}  for the original source extraction. These simulations were made as a function of flux density, creating simulated sources in flux density bins of log$(flux)$ = 0.1, from 300mJy to 1mJy. For each simulation, we injected 100 sources into the original image map and then attempted to recover the sources with SUSSEXtractor. Bright sources in the original maps were masked. Each simulation, at each flux density level, consisted of 100 iterations to ensure a statistically robust measure of the completeness of the source extraction. The completeness correction curves are shown in the right panel of Figure ~\ref{fig:numbercounts}. We find that for the PACS Green 100$\, \mu$m band, the source counts are expected to be 80\% complete at the 8mJy level, dropping to 50\% at $\sim$6mJy. For the PACS Red 160$\, \mu$m band, the 80\% and 50\% completeness levels are reached at 25mJy and 19mJy respectively. Note that blending and flux boosting effects are not thought to be significant given that our counts are well above the PACS 100$\, \mu$m and 160$\, \mu$m confusion limits of 0.75mJy and 3.4mJy respectively (this is also confirmed by our completeness simulations).

\begin{figure*}
 \begin{center}
  \includegraphics[width=8cm]{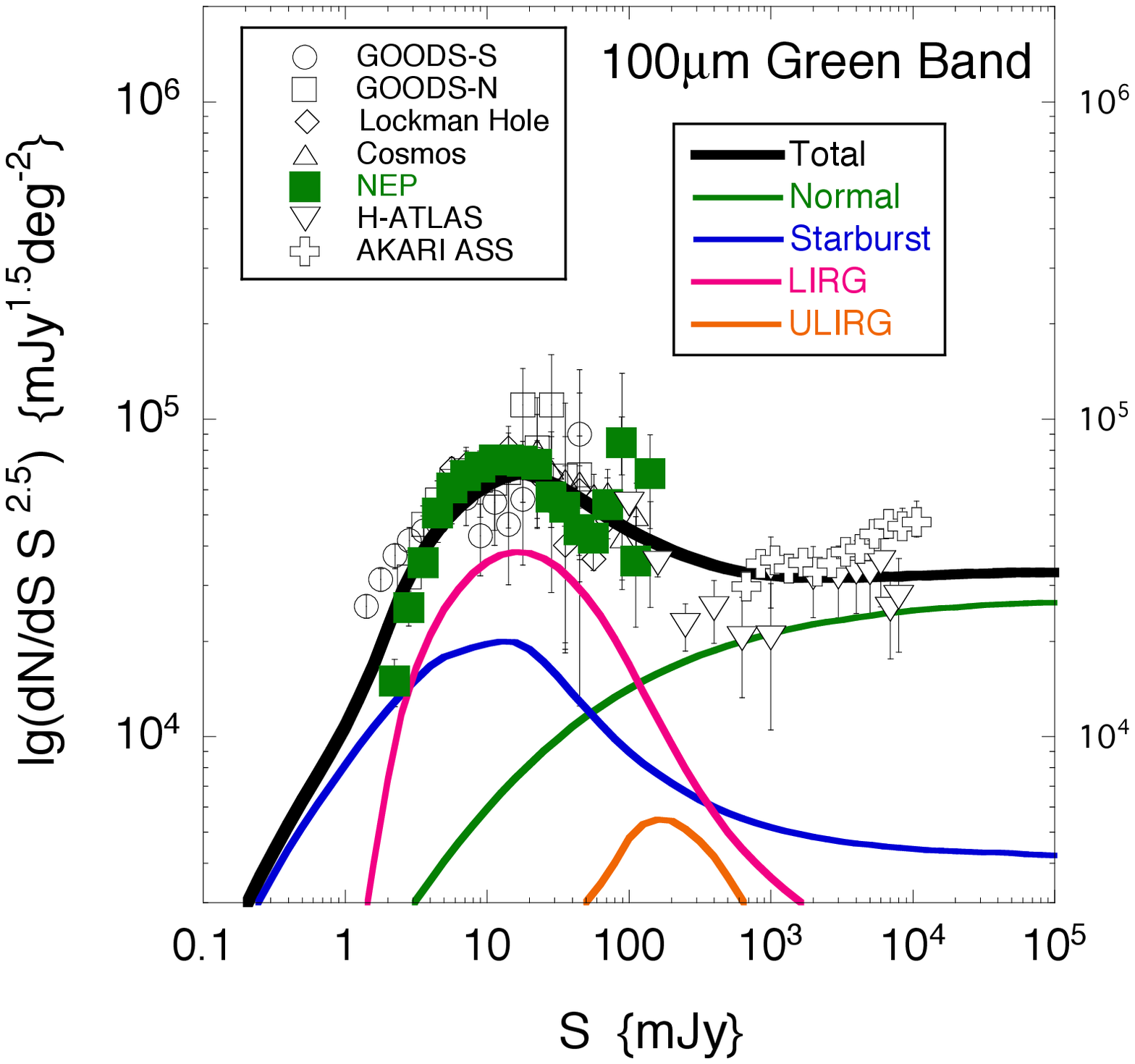} 
  \includegraphics[width=8cm]{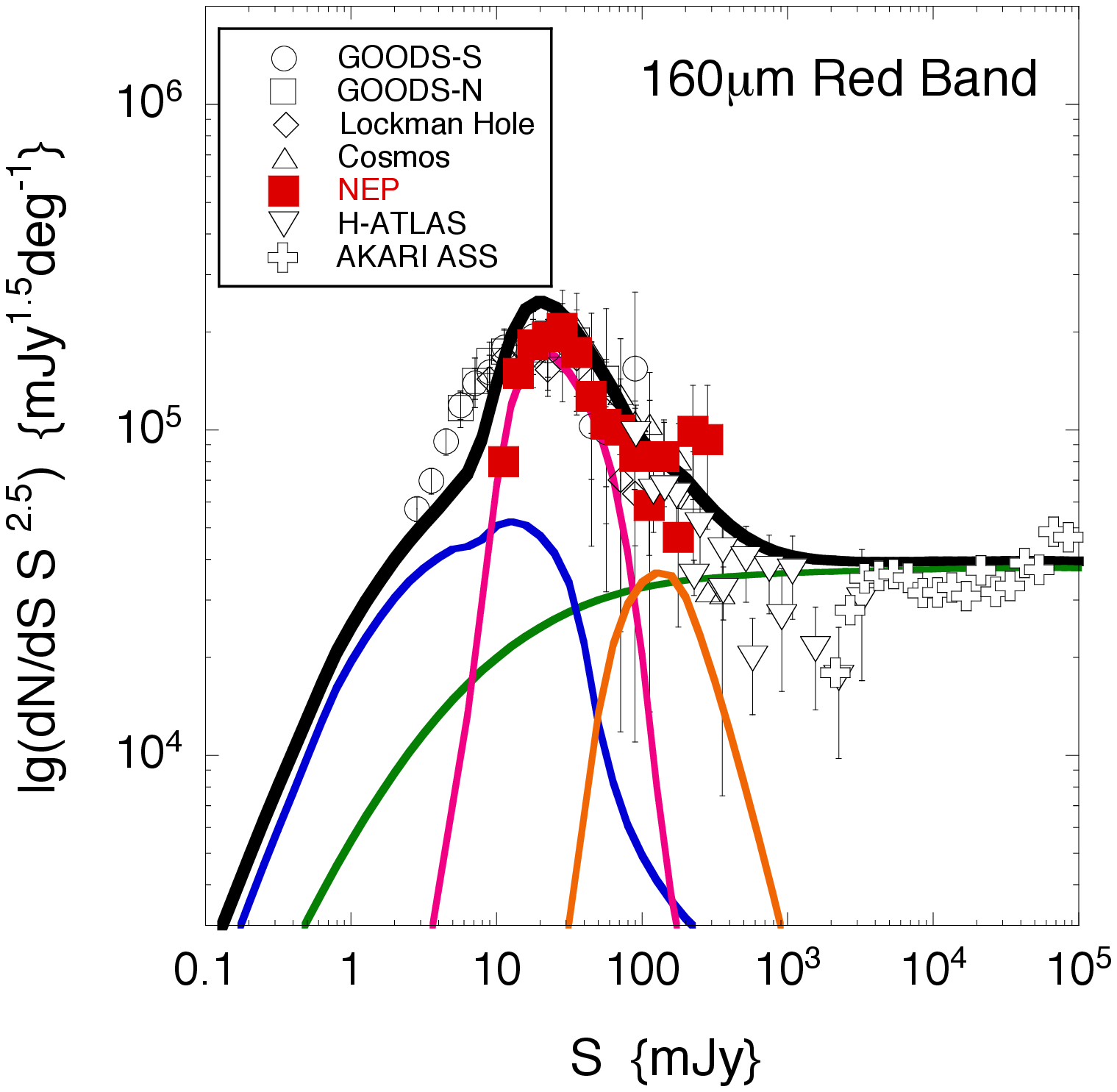} 
 \end{center}
\caption{Final differential source counts normailsed to a Euclidean universe as a function of flux density (mJy). The PACS source counts are shown as  {\it green} and {\it red} squares for the PACS Green 100$\, \mu$m and  Red 160$\, \mu$m bands respectively. Also shown as unfilled symbols, are source counts from other {\it Herschel} surveys and the {\it AKARI} All-Sky survey. The observed counts are compared with the galaxy evolution model of \citet{pearson09, pearson14b} with the thick solid line, the total modelled counts and the individual components for normal (quiescent), starburst $\rm L_{IR}<10^{11}L_\odot$, LIRG $\rm L_{IR}> 10^{11}L_\odot$ and ULIRG $\rm L_{IR}> 10^{12}L_\odot$ components.}
\label{fig:dct}
\end{figure*}

In Figure~\ref{fig:dct}, the final source counts for both PACS bands are plotted as differential number counts (dN/dS S$^{2.5}$) against flux ($S$),  normalised to a Euclidean universe (i.e. N(S) $\propto$ S$^{-3/2}$). The NEP PACS counts are compared with PACS observations from the PEP and H-ATLAS surveys \citep{berta10, berta11, rigby11} and at the brightest fluxes with the {\it AKARI} All-Sky Survey counts reported by \citet{pearson12}, at 90$\, \mu$m  and 160$\, \mu$m ({\it AKARI} WIDE-S and N160 bands, \cite{kawada07}). The NEP counts are in good agreement with PACS observations in other fields over a range of flux densities from 300mJy down to $\sim$10mJy. The faint end of the source counts is not as well constrained since at these levels the estimated completeness is below 50\%.  

\subsection{Comparison with Models}

In Figure~\ref{fig:dct}, we also compare our source counts with a revised galaxy evolution model that extends the evolutionary framework of \citet{pearson09, pearson14b}. This evolutionary model uses the spectral libraries of \citet{efstathiou00, efstathiou03} to model population components including normal quiescent galaxies, star-forming galaxies ($\rm L_{IR}\rm <10^{11}L_\odot$),  luminous (LIRG;  $\rm 10^{11}L_\odot < L_{IR}< 10^{12}L_\odot$)  and ultra-luminous infrared galaxies (ULIRG;  $\rm L_{IR}\rm >10^{12}L_\odot$), via a backwards evolution framework, assuming evolution in both luminosity and number density with redshift (see \citet{pearson09} for details). In the  PACS Green 100$\, \mu$m band, the evolutionary model predictions provide an extremely good fit to the majority of the observed counts, especially over the range in flux density where the evolution in the galaxy population becomes strongly apparent ($S_{100\, \mu m} <$100mJy). 

The models predict that the bulk of the evolutionary contribution to the peak of the source counts arises in the LIRG  $\rm 10^{11}L_\odot < L_{IR}< 10^{12}L_\odot$ population, at redshifts of 0.5$<$z$<$1, consistent with the results of deeper PACS surveys \citep{berta10, magnelli13}. In the longer PACS Red 160$\, \mu$m band, there is also a significant contribution from the more luminous ULIRGs (at z $>$1) at brighter fluxes of 50-100mJy. The bright end of the source counts in both PACS bands is dominated by cool quiescent galaxies. There is some evidence at bright 160$\, \mu$m fluxes that the models may overestimate the counts, however, the H-ATLAS and {\it AKARI} observations are also not in particularly good agreement with each other either.

Using our evolutionary models, we can also estimate the contribution our PACS counts make to the CIRB. At the 50\% completeness level we derive CIRB fractions of $\sim$50$\pm$5\% and $\sim$60$\pm$10\% in the  PACS Green 100$\, \mu$m and Red 160$\, \mu$m bands respectively, compared to the COBE-DIRBE totals of \citet{dole06}, with the errors on the CIRB contribution derived from the deviation of our observations from the evolutionary model. These values are broadly consistent with the values derived from the PEP surveys of 58\% and 74\%, to fainter flux levels of 1.2mJy and 2 mJy respectively \citep{berta11}.


\section{Summary}\label{summary}
A careful re-reduction and analysis of the data from {\it Herschel}/PACS observations at the North Ecliptic Pole (NEP) has been used to produce new image maps in the PACS Green 100$\, \mu$m and Red 160$\, \mu$m bands. These new maps are found to be of superior quality to the current archival Level 2.5 and Level 3 products in the {\it Herschel} Science Archive. A large list of candidate sources derived from these maps was processed through a rigorous quality control and cleaning pipeline to produce  final point source catalogues containing final robust, confirmed source lists of 1384 sources in the Green 100$\, \mu$m band and 630 sources in the Red 160$\, \mu$m band respectively. Comparing these new NEP PACS catalogues with the source lists covering the same area from the {\it Herschel} Science Archive HPPSC, we confirm that the new catalogues reach flux densities more than twice as deep (detecting 400 and 270 more sources in Green 100$\, \mu$m and Red 160$\, \mu$m bands respectively) and provide a more robust set of bright sources than the archival products.

Galaxy source counts derived from the new catalogues are consistent with other PACS surveys over different regions of the sky, confirming the extreme evolution seen in the source counts at fluxes fainter than $\sim$100mJy. Our source counts are estimated to be 50$\%$ complete at the 6mJy and 19mJy level in the Green 100$\, \mu$m and  Red 160$\, \mu$m bands respectively. Comparing our source counts with a model of galaxy evolution, we find that the bulk of the evolving galaxy population is most likely contributed by dusty star-forming galaxies with luminosities $\rm 10^{11}L_\odot < L_{IR}< 10^{12}L_\odot$, i.e. LIRGs, consistent with the results of deeper PACS surveys, and contributing around 50-60$\%$ of the CIRB.

The PACS catalogues at the NEP provide a vital intermediate wavelength ancillary data product, bridging the gap between the {\it AKARI} mid-infrared observations \citep{murata14} and the longer wavelength sub-millimetre observations  with {\it Herschel}-SPIRE (\cite{pearson17}, Pearson et al. in prep.) and the JCMT-SCUBA-2 data \citep{geach17}.

\begin{ack}
The authors would like to thank the anonymous referee for their comments that improved this work. Herschel is an ESA space observatory with science instruments provided by European-led Principal Investigator consortia and with important participation from NASA. HCSS / HSpot / HIPE is a joint development (are joint developments) by the Herschel Science Ground Segment Consortium, consisting of ESA, the NASA Herschel Science Center, and the HIFI, PACS and SPIRE consortia. Campos-Varillas acknowledges support from the FICYT Argo Program. MI acknowledges the support by the grant, No. 2017R1A3A3001362 of the National Research Foundation of Korea (NRF).
\end{ack}


\end{document}